\documentclass[prl,twocolumn,showpacs,floats,floatfix,superscriptaddress]{revtex4-1}

\usepackage{graphicx}
\usepackage{epstopdf}
\usepackage{amsfonts} 
\usepackage{amssymb}
\usepackage{amsmath}
\usepackage{times,mathptmx}

\newcommand{\thetatr}{\theta_\mathrm{tr}}
\newcommand{\thetaml}{\theta_\mathrm{ml}}

\newcommand{\msection}[1]{\vspace{6pt}\paragraph{#1.}}
\newcommand{\tightfig}{\vspace{-12pt}}

\begin{document}

\title{Beyond Fisher: exact sampling distributions of the maximum-likelihood estimator in gravitational-wave parameter estimation}

\author{Michele Vallisneri}
\affiliation{Jet Propulsion Laboratory, California Institute of Technology, Pasadena, CA 91109} 

\begin{abstract}
Gravitational-wave astronomers often wish to characterize the expected parameter-estimation accuracy of future observations. The Fisher matrix provides a lower bound on the spread of the maximum-likelihood estimator across noise realizations, as well as the leading-order width of the posterior probability, but it is limited to high signal strengths often not realized in practice. By contrast, Monte Carlo Bayesian inference provides the full posterior for any signal strength, but it is too expensive to repeat for a representative set of noises. Here I describe an efficient semianalytical technique to map the exact sampling distribution of the maximum-likelihood estimator across noise realizations, for any signal strength. This technique can be applied to any estimation problem for signals in additive Gaussian noise.
\end{abstract}

\maketitle

The first direct detections of gravitational-wave (GW) signals are likely to be achieved in the second half of this decade with second-generation ground-based interferometric detectors such as Advanced LIGO and Virgo \cite{2010CQGra..27h4006H,*avirgo}, which are sensitive to high-frequency GWs (10--1000 Hz). Third-generation instruments such as the Einstein Telescope \cite{grg43} promise much greater reach and yield, while space-based observatories similar to LISA \cite{lisasciencecase} will be sensitive to low-frequency GWs ($10^{-5}$--$10^{-1}$ Hz) in a band populated by thousands of detectable sources in the Galaxy and far beyond. A survey of the scientific literature on GW data analysis in this predetection era reveals a few dominant \emph{genres}: articles on data-analysis methods and implementations \cite[e.g.,][]{2005gr.qc.....9116A}; nondetection and upper-limit analyses of actual collected data \cite[e.g.,][]{1538-4357-683-1-L45};
and ``prospects'' papers that examine combinations of GW sources and detectors to characterize the expected rates of detection and accuracies of source-parameter estimation \cite[e.g.,][]{1994PhRvD..49.2658C}.

Papers in this last class, especially when they are concerned with Bayesian inference for \emph{modeled} GW signals, follow a well-rehearsed structure:
a) derive the GW signal $h$ as a function of the source parameters $\theta$;
b) fix fiducial true values $\theta_\mathrm{tr}$ for them;
c) simulate a realization $n$ of detector noise, usually assumed as additive and Gaussian;
d) derive the noise-dependent probability distribution $p(\theta;n)$ for the source parameters given the data $s = h_\mathrm{tr} + n$;
e) characterize the error and uncertainty of $p(\theta;n)$;
f) finally, repeat steps c, d, and e for a sufficiently broad sample of noise realizations, since each of these will result in different parameter estimates and uncertainties. Optimally, repeat for different fiducial $\thetatr$.

Step d [calculating $p(\theta;n)$] is usually very computationally expensive: for more than two or three source parameters, in Bayesian inference it must be performed with a stochastic technique such as Markov chain Monte Carlo integration \cite{liu2001}, which may involve $\sim 10^6$ evaluations of the likelihood for different $\theta$s. Each evaluation requires obtaining the GW signal $h(\theta)$ and performing an FFT (say, for $10^5$ points) to compute the likelihood of the noise residual [see the discussion around Eq.\ \eqref{eq:mleq}].
Step f is therefore a Monte Carlo integration of Monte Carlos integrations (perhaps again $10^6$ of them), with a total computational cost, for a single $\theta_\mathrm{tr}$, of $\sim 10^{18}$ floating-point operations, more than can usually be procured easily. More emphatically, I call this scheme the Holy Grail of parameter-estimation prospects: even the papers that try hardest \cite[e.g.,][]{2011ApJ...739...99N}, perform meta-Monte Carlo studies of $\sim 100$ combinations of noise realizations and fiducial sources.

Most ``prospects'' papers, instead, take advantage of the fact that steps d to f can be short-circuited when the signal-to-noise ratio (SNR) of detection is sufficiently high that $p(\theta;n)$ collapses to a normal distribution centered around the maximum-likelihood parameters $\theta_\mathrm{ml}(n)$, with covariance given by the inverse Fisher matrix, which is \emph{not} a function of $n$. At high SNR, the distribution of $\theta_\mathrm{ml}$ for different noise realizations is itself normal and centered at $\theta_\mathrm{tr}$, with covariance again given by the inverse Fisher matrix. The computational cost of this approximation (for a single $\theta_\mathrm{tr}$ and $10^5$-point likelihoods) is $\sim d^2 \times 10^6$ floating-point operations, where $d$ is the number of source parameters, affording virtually instantaneous results.

Unfortunately, as I discuss at length in Ref.\ \cite{Vallisneri2008fisher} and as highlighted elsewhere \cite{1994PhRvD..49.2658C,PhysRevD.53.3033,*PhysRevD.57.3408,*PhysRevD.57.4588,*PhysRevD.82.124065}, for many practical parameter-estimation problems SNRs will not be sufficiently high to justify the approximation. The crux of the problem is that the Fisher matrix, built from the partial derivatives $\partial h/\partial \theta_i$ of the signal, can represent $h$ correctly only if $h$ is \emph{linear} in all the $\theta_i$ across ranges comparable to the expected parameter errors. These decrease as the SNR grows, making the condition less stringent; in Ref.\ \cite{Vallisneri2008fisher}, I provide a criterion to determine when the SNR is high enough.

\begin{figure}
\includegraphics[width=\columnwidth]{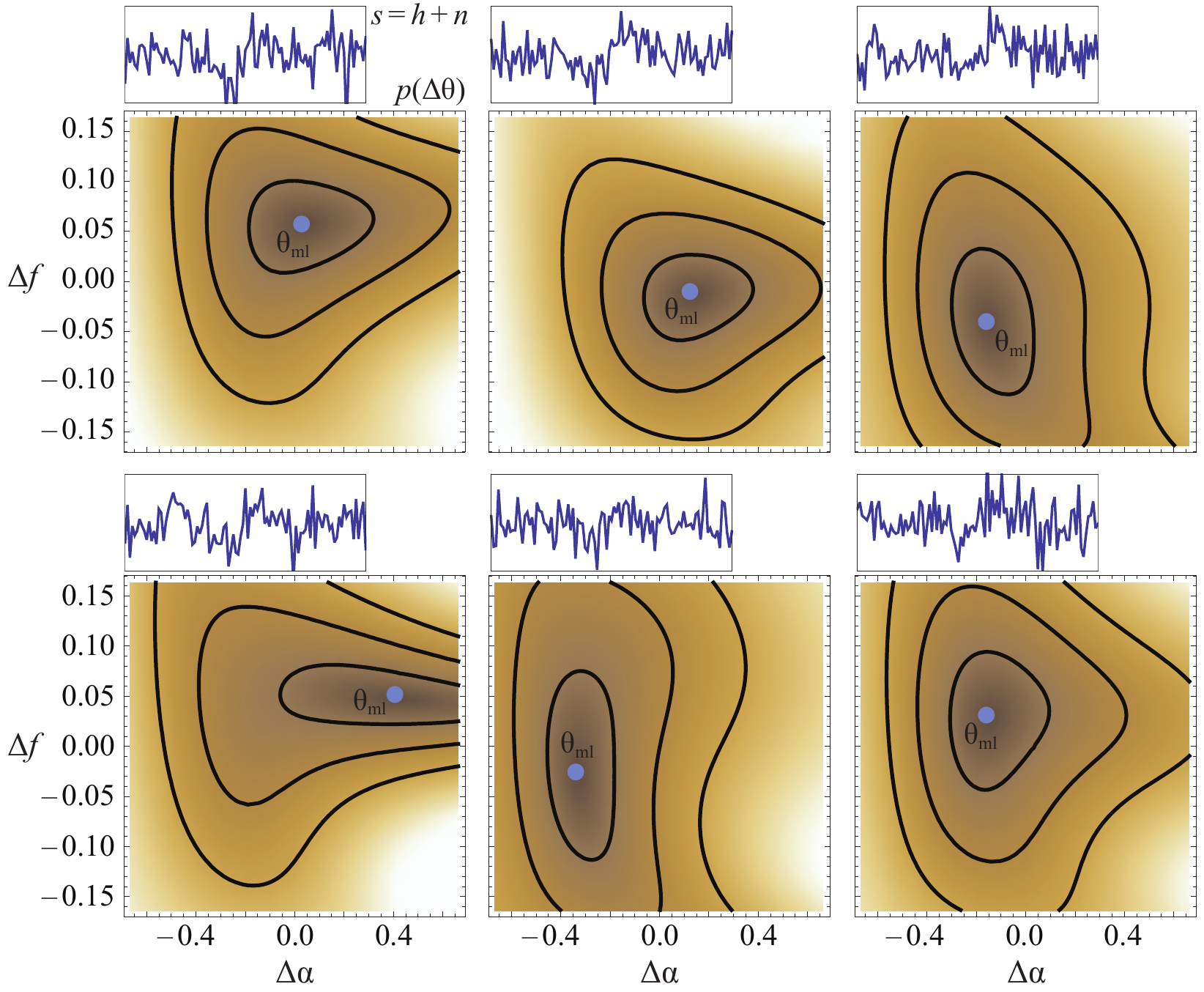}\vspace{-4pt}
\caption{Likelihood maps obtained for the toy model \eqref{eq:toy} with different $n$'s and with the same ($\alpha_\mathrm{tr},f_\mathrm{tr}$), located at the origin of the plots.
Contours are plotted at levels that would correspond to 1-$\sigma$, 2-$\sigma$, and 3-$\sigma$ ellipses for a normal distribution. The signal $s = h + n$ is shown above each map.
\label{fig:like}}
\tightfig
\end{figure}

Thus, in general a Monte Carlo integration is needed for step d; furthermore, the meta-Monte Carlo integration of step f is also necessary, because the particular noise realization \emph{does} affect parameter uncertainties, as shown in Fig.\ \ref{fig:like} for the toy model discussed later in this Letter.
It is true \cite{2010ApJ...725..496N} that for additive Gaussian noise the average of $p(\theta;n)$ across all $n$'s equals $p(\theta;n=0)$; but this averaged likelihood describes an unrealistic limit (infinite observations of the same source with a finite total SNR) that is not representative of average or typical errors, except trivially in the high-SNR limit. Thus, the shape and width of generic $p(\theta;n)$ cannot generally be obtained from $p(\theta;n=0)$, as suggested in Ref.\ \cite{2010arXiv1007.4820C}. In Ref.\ \cite{Vallisneri2008fisher} I derive an expansion of $p(\theta;n)$ in powers of $1/\mathrm{SNR}$, where $n$-dependent terms are seen at next-to-leading order for \emph{both} the centering and shape of the posterior.

In this Letter I describe and test a technique to perform as less thorough, but more affordable survey than a nested Monte Carlo: mapping the sampling distribution of the maximum-likelihood (henceforth, ML) estimator $\theta_\mathrm{ml}$ [the maximum of each $p(\theta;n)$] over all noise realizations. From a classical-statistics (a.k.a.\ ``frequentist'') viewpoint, the spread of this distribution is formally the uncertainty of the ML point estimator. From a Bayesian viewpoint \cite{Jaynes2003}, if priors are unimportant, the spread of this distribution describes one major component of the expected error (the other is the intrinsic spread of the posterior for each $n$).

\msection{Mapping the sampling distribution of the ML estimator}

We consider a large set of experiments in which we observe the signal $h(\theta_\mathrm{tr})$, where $\theta^i_\mathrm{tr}$ is the $d$-dimensional vector of true parameter values and $h$ is an $N$-dimensional vector (e.g., a time series) with $N \gg d$. In each experiment, the detector output is $s = n + h(\theta_\mathrm{tr})$, where $n$ is additive Gaussian noise distributed with $p(n) = \mathcal{N} \exp -(n,n)/2$. Here $(\cdot,\cdot)$ is the standard signal inner product, given in one convention \cite{1994PhRvD..49.2658C} by
$(s,t) = 4 \, \mathrm{Re} \int_0^\infty \tilde{s}^*(f) \tilde{t}(f) / S(f) \, \mathrm{d}f$, with $\,\tilde{}\,$ denoting the Fourier transform and ${}^*$ the complex conjugate. In this Letter, without loss of generality, we treat $(\cdot,\cdot)$ as the inner product of an abstract linear space, and let $|s|^2 \equiv (s,s)$.

For a given $\theta_\mathrm{tr}$ and noise realization $n_\mathrm{tr}$,
the ML estimator $\theta_\mathrm{ml}(n_\mathrm{tr},\theta_\mathrm{tr})$ maximizes the probability of the residual noise $n$ obtained by subtracting the postulated signal $h(\theta)$ from the data $s = h(\theta_\mathrm{tr}) + n_\mathrm{tr}$; that is, it maximizes
\begin{equation}
p(n = s - h(\theta)) \propto \mathrm{e}^{-|h(\thetatr) + n_\mathrm{tr} - h(\theta)|^2/2}.
\label{eq:mleq}
\end{equation}
Thus, $\theta_\mathrm{ml}$ must satisfy the vector equation
\begin{equation}
\mathrm{ML}_i(\theta;n,\thetatr) \equiv (\partial_i h(\theta),h(\thetatr) + n - h(\theta)) = 0,
\label{eq:mlsimp}
\end{equation}
where the $\mathrm{ML}_i$ are the partial derivatives of $-2 \log p$.
%

Our purpose is to map the distribution of $\thetaml$ across all noise realizations.
We can do this by enumerating all possible $n$'s [weighted by $p(n)$], figuring out the $\thetaml$ corresponding to each, and accumulating the resulting distribution of $\thetaml$. Formally,
\begin{equation}
p(\thetaml=\theta|\thetatr) = \int \delta\bigl(\thetaml(n,\thetatr) - \theta\bigr) \, p(n) \, \mathrm{d}n.
\label{eq:dirint}
\end{equation}
Unfortunately, because $h(\theta)$ is generally a complicated function of the $\theta^i$, it is difficult to solve $\mathrm{ML}_i = 0$
for $\theta_\mathrm{ml}$ given $n$, so we can only integrate Eq.\ \eqref{eq:dirint} by using a Monte Carlo approach where we generate full, high-dimensional realizations of $n$ (e.g., as time series), search parameter space for $\thetaml$ by repeatedly evaluating Eq.\ \eqref{eq:mleq}, and iterate for many different $n$.

The main result of this paper is that there \emph{is} a more effective way to map  $p(\thetaml)$: we enumerate the $\thetaml$, and compute the total probability weight of the $n$ that are compatible with each. (Put slightly differently, for each $\thetaml$ we count how many experiments would yield it as the maximum of likelihood.) Because the $\mathrm{ML}_i$ are \emph{linear} functions of the $n$, this weight is the probability mass of the $(N-d)$-dimensional subspace of the noise vectors that solve $\mathrm{ML}_i = 0$ for all $i$. Using the $\delta$-of-function relation for the vector $\mathrm{ML}_k$, $\delta(\mathrm{ML}_k(\theta;n,\thetatr)) = \delta(\thetaml(n,\thetatr) - \theta) / |\partial \mathrm{ML}_i(\theta)/\partial \theta_j|$, we can then rewrite $p(\thetaml = \theta|\thetatr)$ as
\begin{equation}
\mathcal{N} \int
\Pi_k \delta\big(\mathrm{ML}_k(\theta;n,\thetatr)\big) \times \left| \partial \mathrm{ML}_i/\partial \theta_j \right| \mathrm{e}^{-(n,n)/2} \, \mathrm{d} n.
\label{eq:master}
\end{equation}
\vspace{-12pt}
\begin{figure}
\includegraphics[width=\columnwidth]{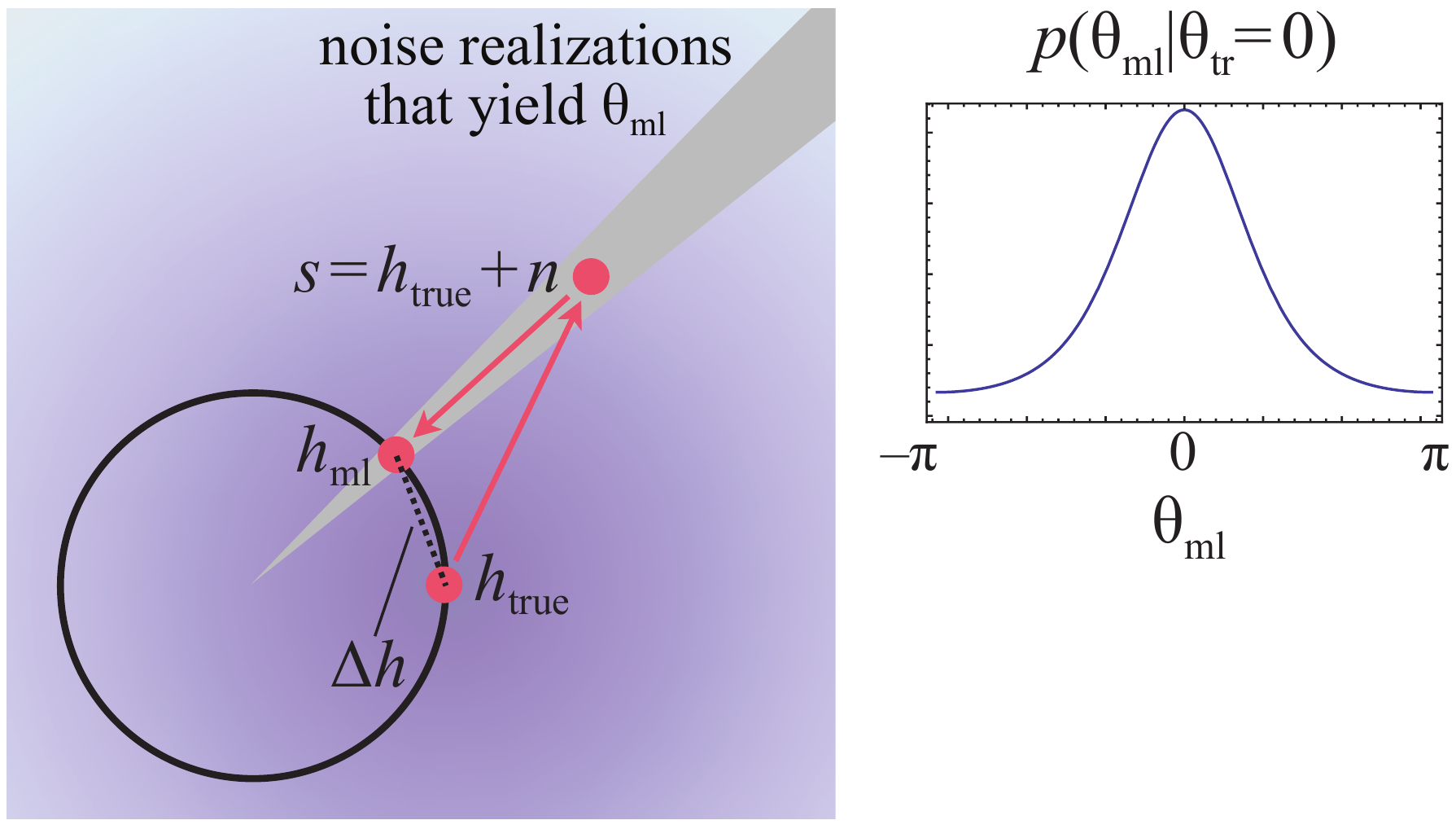}\vspace{-4pt}
\caption{
To compute $p(\thetaml)$, we integrate $p(n)$ over all noise realizations that satisfy $\mathrm{ML}_i = 0$---see the discussion below Eq.\ \eqref{eq:master}. Here $h(\theta) = (\cos \theta,\sin \theta)$, $p(n_x,n_y) = (2 \pi)^{-1} \exp -(n_x^2 + n_y^2)/2$, and $\thetaml = \arctan (s_y/s_x)$. The resulting 
$p(\thetaml|\thetatr=0)$ is shown top right.
\label{fig:proc}}
\tightfig
\end{figure}

Figure \ref{fig:proc} exemplifies the process of integrating over the $n$'s compatible with a chosen $\theta_\mathrm{ml}$. In this low-dimensional model \footnote{Suggested by C.\ Cutler in a personal communication (2010).}, detector data are described by a point in the plane, and the signal family $h(\theta)$ by the unit circle.
For each $n$, the true signal $h_\mathrm{true}$ is displaced to a different $s = h_\mathrm{true} + n$ with probability $p(n)$ indicated by the shading. Projecting back to $h(\theta)$ 
identifies the ML waveform $h_\mathrm{ml}$ and parameters $\theta_\mathrm{ml}$. All noise realizations that produce an $s$ within the gray sector project to the same $h_\mathrm{ml}$, so integrating $p(n)$ over the sector yields $p(\theta_\mathrm{ml})$.

Equation \eqref{eq:master} is especially powerful because, contrary to appearance, it does not require integration over the full $N$ dimensions of the noise, but only over $\sim d^2$ coordinate directions corresponding to the $\mathrm{ML}_k$ and $\partial \mathrm{ML}_i/\partial \theta_j$. Since both these sets of functions are linear in $n$, they can be seen as jointly normal random variables that are fully characterized by their inner products; all the other noise degrees of freedom have no effect on the integral other than its normalization. (We spell this out in detail in the next section.)
With Eq.\ \eqref{eq:master} in hand, we can then sample $p(\theta_\mathrm{ml})$ directly (for low $d$), or by Markov chain Monte Carlo techniques.

\msection{Evaluating the master integral}
\label{sec:integ}

Equation \eqref{eq:master} can be evaluated elegantly as the expectation value of a function (the determinant) of correlated random variables. For clarity, we follow a more pedestrian approach: we begin by transforming the $N$-dimensional integral over the $n$ to new coordinates where the first few basis vectors span the random variables of interest. Namely, we write $n$ in terms of a new basis where the first $d$ vectors are obtained by orthonormalizing the $\partial_i h \equiv h_i$,
\begin{equation}
\hat{n}_1 \propto h_1, \quad
\hat{n}_i \propto \Bigl(1 - \sum_{j=1}^{i-1}\hat{n}_j \otimes \hat{n}_j\Bigr) \cdot h_i;
\end{equation}
%
furthermore, we obtain the $(d+1)$-th to $d(d+3)/2$-th basis vectors by orthonormalizing the $\partial_{ij} h \equiv h_{i,j}$ with $i \geq j$; the remaining $N - d(d+3)/2$ noise combinations complete the basis.

We can now rewrite all the variables that appear in Eq.\ \eqref{eq:master} in terms of the new basis:
\begin{equation}
n \equiv N^k \hat{n}_k, \quad h_i \equiv C_i^k \hat{n}_k, \quad h_{i,j} \equiv C_{ij}^k \hat{n}_k;
\end{equation}
here the $N^k$, $C_i^k$, and $C_{ij}^k$ are the coefficients that expand the $n$, $h_i$, and $h_{i,j}$, respectively, in terms of the $\hat{n}_k$.
We use Einstein summations, and treat $(ij)$ both as a double index and as a single index flattened to the range $(d+1),\ldots,d(d+3)/2$.
By virtue of orthonormalization, $C_i^k = 0$ for $k > i$ and $C_{ij}^k = 0$ for $k > (ij)$, with $C_i^i > 0$ and $C_{ij}^{(ij)} > 0$.
Furthermore,
\begin{equation}
\begin{aligned}
\mathrm{ML}_i &\equiv C_i^k N_k - C_i^k (\Delta h,\hat{n}_k), \\
\mathrm{ML}_{i,j} &\equiv C_{ij}^m N_m - C_{ij}^m (\Delta h,\hat{n}_m) - C_i^k C_{jk},
\end{aligned}
\end{equation}
with $\Delta h \equiv h(\theta) - h(\theta_\mathrm{tr})$, and with the sums over $k$ limited to $k \leq i$ and the sums over $m$ limited to $m \leq (ij)$.

To perform the integration in Eq.\ \eqref{eq:master}, we use the first $\delta$ to fix $N_1 = (\Delta h,\hat{n}_1)$, yielding a $\delta$-normalization factor of $1/C_1^1$; we use the second $\delta$ to fix $N_2 = (\Delta h,\hat{n}_2)$ (after the terms proportional to $N_1$ in the $\delta$ cancel out), yielding a factor of $1/C_2^2$; and so on. Next, we perform the trivial integral over the ``signal-orthogonal'' noise degrees of freedom that correspond to the $\hat{n}_k$ with $k > d(d+3)/2$, leaving
\begin{multline}
p(\thetaml = \theta|\thetatr) = \frac{e^{
-\Sigma_{k=1}^d (\Delta h,\hat{n}_k)^2 / 2
}}{(2\pi)^{d(d+3)/4} \Pi_{k=1}^d C_k^k} \times \\
\int
\left| C_{ij}^m N_m - C_{ij}^m (\Delta h,\hat{n}_m) - C_i^k C_{jk} \right|
e^{
-N^m N_m / 2
} \,
\mathrm{d}N_m;
\label{eq:intnum}
\end{multline}
here the sums over $m$ span \footnote{Inside the determinant, the first $d$ terms of the form $C_{ij}^m [N_m - (\Delta h,\hat{n}_m)]$ vanish because of the $\delta$s applied previously.} $m = (d+1),\ldots,d(d+3)/2$, and the sum over $k$ spans $k = 1,\ldots,d$.

For a given $\theta$, the main computational cost of evaluating Eq.\ \eqref{eq:intnum} resides in the orthonormalization and in the computation of the $(\Delta h,\hat{n}_m)$, which together require $\sim d^4/8$ inner products (and therefore $N$-dimensional signal FFTs); by contrast, the $d(d+1)/2$-dimensional Gaussian integral can be evaluated much more cheaply (e.g., by a 10,000-point Monte Carlo integration over $N_m$ drawn from a normal distribution), since the integrand is a function of small matrices and not long FFTs.

\msection{Fisher-matrix limit}

The well-known high-SNR, Fisher-matrix limit, in which the waveform can be approximated as a linear function of the parameters (and in which the $\theta^i_\mathrm{ml}$ have a simple jointly-normal distribution), follows easily by specializing Eq.\ \eqref{eq:intnum}. Without loss of generality, we set $h(\theta) = \theta^i h_i$;
it follows that $h_{i,j} = C^k_{ij} = 0$. The integration over the $N_m$ is trivial, and yields $(2\pi)^{d(d-1)/4} |C_i^k C_{jk}|$, where $C_i^k C_{jk} = (h_i,h_j) \equiv F_{ij}$ is the Fisher matrix.
Furthermore, $\Delta h = h_i \Delta \theta^i$, where $\Delta \theta^i \equiv \theta_\mathrm{ml}^i - \theta^i_\mathrm{tr}$ is the \emph{error} of the ML estimator, so the first exponential of Eq.\ \eqref{eq:intnum} can be rewritten as $\exp(-\Delta \theta^i F_{ij} \Delta \theta^j/2)$, since the $\hat{n}_k$, for $k = 1,\ldots,d$, span a complete basis for the $h_i$. Last, because of the structure of the $C_i^k$, $\Pi_{k=1}^d C_k^k = \sqrt{|C_i^k C_{jk}|} = \sqrt{|F_{ij}|}$. Taking everything together, we reproduce the Fisher-matrix result for the distribution of $\thetaml$ \footnote{Formally, we see that Eq.\ \eqref{eq:intnum} tends to Eq.\ \eqref{eq:fisherlimit} because the terms other than $F_{ij}$ in the determinant are suppressed by a factor 1/SNR. This follows from $C_i^k$ and $C_{ij}^m \sim O(\mathrm{SNR})$, $N_m$ and $\hat{n}_m \sim O(1)$, and $\Delta h \sim O(\mathrm{SNR}) \times O(\sqrt{F^{-1}}) \sim O(1)$.},
\begin{equation}
\label{eq:fisherlimit}
p(\thetaml|\thetatr) = \frac{e^{- 
\Delta \theta^i F_{ij} \Delta \theta^j/2}}{\sqrt{(2\pi)^d |F^{-1}_{ij}|}}.
\end{equation}

The general result [Eq.\ \eqref{eq:intnum}] can be restated in terms of $F_{ij}$, in a form more suitable to computation:
\begin{multline}
\label{eq:intnum2}
p(\thetaml = \theta|\thetatr) =
\frac{e^{
-(\Delta h,h_i) (F^{-1})^{ij} (\Delta h,h_j) / 2
}}{\sqrt{(2\pi)^d |F_{ij}|} \sqrt{(2\pi)^{d(d-1)/2} |D_{\mu\nu}|}} \times \\
\int
\left|F_{ij} + (\Delta h,h_{ij}) - M_{(ij)} \right|
e^{-M_\mu (D^{-1})^{\mu \nu} M_\nu / 2}
\, \mathrm{d}M_\mu;
\end{multline}
here $D_{\mu\nu} \equiv D_{(ij)(kl)}$ is the $d(d-1)/2$-dimensional square matrix given by the products $(h'_{i,j},h'_{k,l})$ with $j\leq i$, $l \leq k$: the primes denote projection orthogonal to $h_k$ (i.e., $h'_{i,j} = \sum_{d+1}^{d(d+3)/2} C^k_{ij} \hat{n}_k$);
and $M_{(ij)}$ is the matrix obtained from the $d(d-1)/2$-dimensional vector
$M_\mu$ of integration variables by remapping indices.

\msection{Toy model}
To exemplify the use of Eqs.\ \eqref{eq:master} to map $p(\thetaml)$ for low-SNR parameter estimation, let us consider a family of sine--Gaussian signals given by 
\begin{equation}
h(t;A,\alpha,f) = A \, \mathrm{e}^{-t^2/2\alpha^2} \sin(2\pi f t).
\label{eq:toy}
\end{equation}
We consider the problem of jointly estimating $\alpha$ and $f$, but not $A$, which
we fix to yield SNR = 5. For our example, we select $\alpha_\mathrm{tr} = 1$ and $f_\mathrm{tr} = 0.25$. 
%
As shown in Fig.\ \ref{fig:like}, at this low SNR different noise realizations yield strikingly different likelihood maps for the same true signal---all quite different from the ellipsoidal Fisher prediction. In each map, the pale blue dot marks the location of $\theta_\mathrm{ml}$---indeed, the distribution of blue dots is just the desired $p(\thetaml)$.

\begin{figure}
\includegraphics[width=\columnwidth]{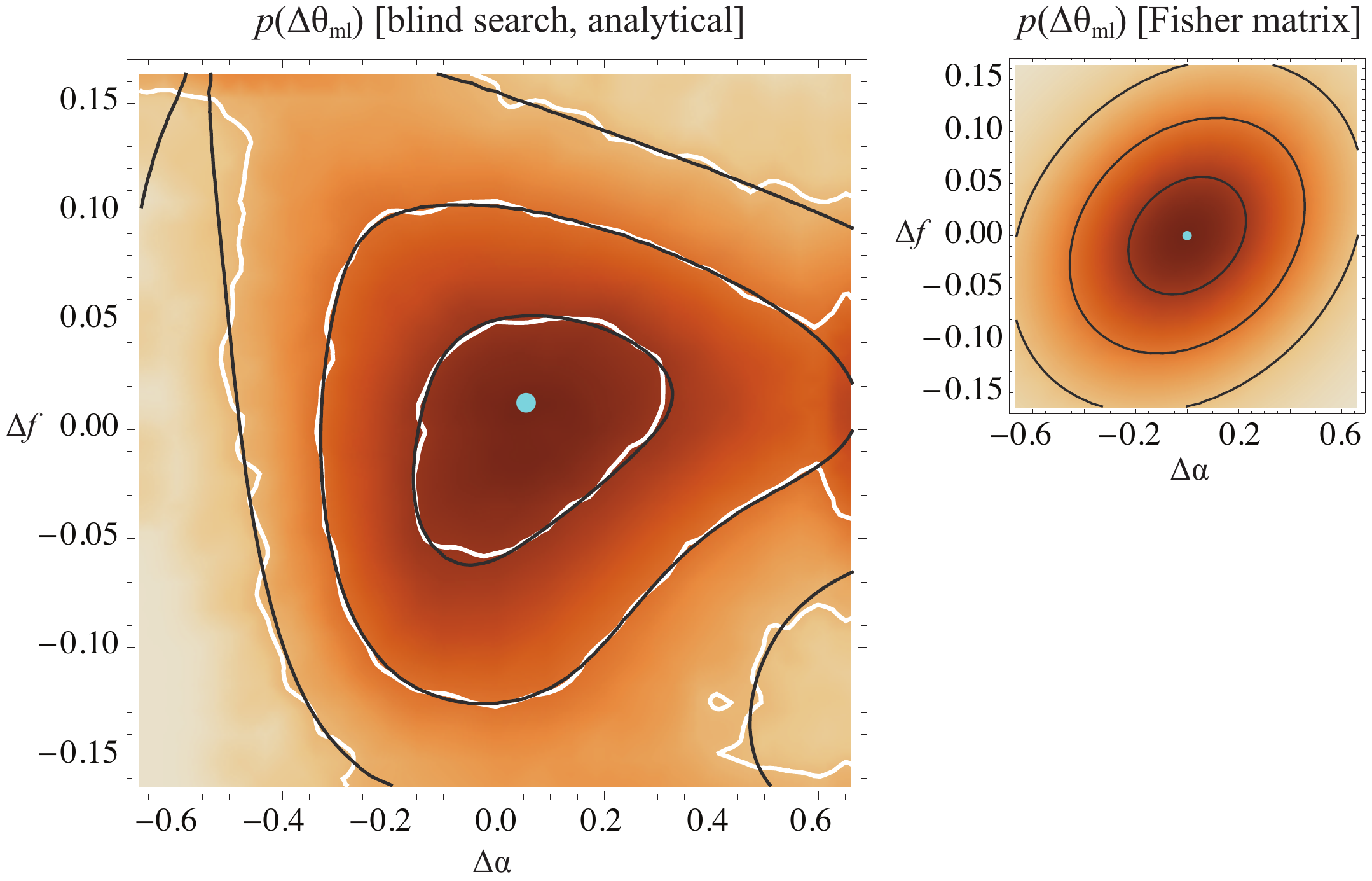}\vspace{-4pt}
\caption{Exact (left) and Fisher-matrix (right) sampling distribution of $\thetaml$. Contours are plotted as in Fig.\ \ref{fig:like}, with the noisy white curves and shading derived
from the numerical maxima of 100,000 likelihood maps for different $n$'s,
and the dark curves from Eq.\ \eqref{eq:master}.
The dot marks the most probable $\theta_\mathrm{ml}$, slightly displaced from $\theta_\mathrm{tr}$.
\label{fig:map}}
\tightfig
\end{figure}

In Fig.\ \ref{fig:map}, I show maps of $p(\thetaml)$ obtained by using both the brute-force approach [Eq.\ \eqref{eq:dirint}] and the new method [Eq.\ \eqref{eq:master}].
For the former (the noisy white curves and shading), I produced 100,000 likelihood maps $p(\theta;n)$ with different $n$'s drawn from $p(n)$, and 2D-histogrammed the resulting $\theta_\mathrm{ml}$. For the latter (dark curves), I used Eq.\ \eqref{eq:master}, as implemented by Eq.\ \eqref{eq:intnum}.
As expected, the maps agree, but the new method is considerably faster. For comparison, the top--right plot shows the Fisher-matrix prediction \eqref{eq:fisherlimit}.

\msection{Conclusions}

I have described a novel approach to create exact maps, for any SNR, of the distribution of the ML estimator for the source parameters of a signal embedded in additive Gaussian noise. This distribution would be obtained in a large set of observations of the same true signal with different noise realizations, each appearing with probability $p(n)$. Given a single observation, such a map embodies the frequentist notion of uncertainty for the ML estimator. From a Bayesian viewpoint, if priors are unimportant, the map characterizes the distribution of possible maxima of posterior probabilities.

In comparison to the computational cost of the ``Holy Grail'' nested Monte Carlo ($10^{18}$ operations), we estimate the cost of Eq.\ \eqref{eq:master} as follows: $10^6$ Monte Carlo samples of candidate $\theta_\mathrm{ml}$s, times $d^4/8$ inner products, times $10^6$ floating-point operations for each inner-product FFT (again assuming $N = 10^5$); thus, even for $d = 10$, this scheme involves $\sim 10^{15}$ operations---a thousand times cheaper.

These maps can be used directly, in both frequentist and Bayesian frameworks, to study parameter-estimation prospects, but also to perform stringent tests of Fisher-matrix predictions at low SNRs, and to provide proposal distributions for Monte Carlo searches of unknown sources. An interesting feature in this regard is that $\mathrm{ML}_i = 0$,
a local condition, does not distinguish between primary and secondary maxima of the likelihood, and it will include the latter (if they are sufficiently probable) in the maps; thus Eq.\ \eqref{eq:master} could be exploited to enable jumps between separated peaks in complex likelihoods, which are generally very difficult to locate.
While this result was derived in the context and with the motivation of GW science, it is applicable to statistical inference for any problem where noise can be regarded as Gaussian and additive, such as several that arise in high-energy physics and observational cosmology.

\msection{Acknowledgments}
I am grateful to G.\ Cicuta, N.\ Cornish, C.\ Cutler, F.\ Feroz, M.\ Hobson, J.\ Jewell, I.\ Mandel, S.\ Nissanke, E.\ Onofri, R.\ O'Shaughnessy, and T.\ Prince, as well as two anonymous referees, for useful suggestions and for reviewing this manuscript. This work was supported by the RTD program at the Jet Propulsion Laboratory, California Institute of Technology, where it was performed under contract with the National Aeronautics and Space Administration. Copyright 2011 California Institute of Technology. Government sponsorship acknowledged.

\bibliography{gene}

\end{document}